\documentclass[fleqn,usenatbib]{mnras}

\usepackage{newtxtext,newtxmath}

\usepackage[T1]{fontenc}
\usepackage{ae,aecompl}

\usepackage{graphicx}	
\usepackage{amsmath}	
\usepackage{amssymb}	
\usepackage{bm}		
\usepackage{hyperref}

\title[Unbound stars and predictions for \textit{Gaia}]{Dynamical evolution of star-forming regions: III. Unbound stars and predictions for \textit{Gaia}}

\author[Schoettler et al.]{Christina Schoettler,$^{1,2}$\thanks{E-mail: cschoettler1@sheffield.ac.uk (CS)}
Richard J. Parker,$^{1}$\thanks{Royal Society Dorothy Hodgkin Fellow (RJP)}
Becky Arnold,$^{1}$ 
Liam P. Grimmett,$^{1}$ 
\newauthor
Jos de Bruijne,$^{2}$ Nicholas J. Wright$^{3}$ 
\\
\\
$^{1}$Department of Physics and Astronomy, The University of Sheffield, Hicks Building, Hounsfield Road, Sheffield S3 7RH, UK\\
$^{2}$Science Support Office, Directorate of Science, European Space Research and Technology Centre (ESA/ESTEC),\\ 
Keplerlaan 1, NL-2201 AZ Noordwijk, the Netherlands\\
$^{3}$Astrophysics Group, Keele University, Keele ST5 5BG, UK}

\date{Accepted XXX. Received YYY; in original form ZZZ}

\pubyear{2019}

\begin{document}
\label{firstpage}
\pagerange{\pageref{firstpage}--\pageref{lastpage}}
\maketitle

\begin{abstract}
We use $N$-body simulations to probe the early phases of the dynamical evolution of star-forming regions and focus on mass and velocity distributions of unbound stars. In this parameter space study, we vary the initial virial ratio and degree of spatial and kinematic substructure and analyse the fraction of stars that become unbound in two different mass classes (above and below 8 M$_{\sun}$). We find that the fraction of unbound stars differs depending on the initial conditions. After 10 Myr, in initially highly subvirial, substructured simulations, the high-mass and lower-mass unbound fractions are similar at $\sim$23 per cent. In initially virialised, substructured simulations, we find only $\sim$16 per cent of all high-mass stars are unbound, whereas $\sim$37 per cent of all lower-mass stars are. The velocity distributions of unbound stars only show differences for extremely different initial conditions. The distributions are dominated by large numbers of lower-mass stars becoming unbound just above the escape velocity of $\sim$3\,km\,s$^{-1}$ with unbound high-mass stars moving faster on average than lower-mass unbound stars. We see no high-mass runaway stars (velocity > 30\,km\,s$^{-1}$) from any of our initial conditions and only an occasional lower-mass runaway star from initially subvirial/substructured simulations. In our simulations, we find a small number of lower-mass walkaway stars (with velocity 5-30\,km\,s$^{-1}$) from all of our initial conditions. These walkaway stars should be observable around many nearby star-forming regions with \textit{Gaia}.


\end{abstract}




\section{Introduction}

The majority of stars do not form in isolation but in environments where stellar densities are higher than in the Galactic field \citep{RN25, RN59}. Depending on the initial conditions in these star-forming regions, they either evolve into bound clusters or unbound groups. Most stars born in clusters do not remain there past an age of 10 million years (Myr) and only about 10 per cent are observed in gravitationally bound clusters at this age \citep{RN25}. Young stars that are not members of bound clusters are usually observed in unbound groups of stars \citep[i.e. associations,][]{RN19} before they disperse into the Galactic field. 

For the last two decades, the prevailing view has been that bound star clusters are fundamental units of star formation - in that most stars form in these dense, embedded environments until gas exhaustion \citep{RN292} or residual gas expulsion conclude star formation. Gas expulsion can also lead to the cluster's dissolution \citep{RN283,RN284,RN285,RN49,RN292,RN286}. In this scenario, associations must have formed as single or multiple clusters and expanded into their unbound state \citep[so-called monolithic star formation - e.g.][]{RN268,RN274}. An alternative view is that depending on the initial conditions of the molecular clouds, clusters or associations are formed when smaller clustered regions with differing stellar densities assemble hierarchically. These smaller groups of stars are the result of hierarchical fragmentation of the molecular clouds. In this scenario, star formation can lead to the formation of a dense, bound star cluster but can also result in lower-density, unbound associations \citep[so-called hierarchical star formation - e.g.][]{RN275,RN267}. Recent work has shown that associations are unlikely to be dissolved clusters, supporting the latter star formation scenario \citep[e.g.][]{RN6, RN80, RN186}.

One way of testing these two scenarios observationally is to determine the initial density (i.e. spatial structure) and virial ratio (i.e. velocity structure) of star-forming regions. This remains difficult as dynamical evolution can lead to significant changes to star-forming regions over a short period of time \citep[e.g.][]{RN4, RN258} such as a rapid reduction in density in regions with initially high stellar densities \citep[e.g.][]{RN277,RN8}. Initial substructure can be erased \citep[e.g.][]{RN14,RN4,RN248}, the location of the most massive stars can change due to dynamical mass segregation \citep[e.g][]{RN15,RN5} and primordial binary systems can be destroyed \citep[e.g.][]{RN261,RN256,RN277,RN257}. A dense phase can also have disruptive effects on protoplanetary discs and young planetary systems around stars in star-forming regions \citep[e.g.][]{RN271,RN272,RN269,RN270,2019arXiv190211094N}.

Our earlier work showed that information from the spatial distribution of star-forming regions can be used to distinguish the initial bound/unbound state (initial virial ratio) \citep[][Paper I]{RN5}. \citet[][Paper II]{RN1} showed that using radial velocity dispersion in combination with a spatial structure diagnostic \citep[Q-parameter,][]{RN16} can help constrain initial conditions in star-forming regions with high local densities. In this paper, we will focus on stars that become unbound from young star-forming regions. 

Observationally, unbound stars are easiest to identify when their velocities are higher than their surroundings and they have high mass and therefore high luminosity, such as OB stars. These stars have a mass of at least 8 M$_{\sun}$ and short lifetimes of up to a few tens Myr, with the most massive stars undergoing core-collapse supernovae after only a few Myr \citep{RN10}. Most star-forming regions dissolve after only a few tens Myr but they can still outlive the massive stars located within them \citep{RN21}. As a consequence, OB stars should not be found outside these regions. However, there are OB stars found far outside star-forming regions moving at much higher velocities than normally expected. These OB stars have been termed runaway stars by \citet{RN67}. They show a peculiar velocity (i.e. velocity relative to a rest frame) in excess of $\sim$30-40\,km\,s$^{-1}$ and/or are located at large distances from any star-forming regions or the Galactic plane \citep[e.g.][]{RN255, RN67, RN276, RN50, RN190, RN137,RN293}. Almost all currently identified runaway stars are high-mass stars \citep[see the recent catalogue in][]{RN263}. Low-mass runaway star detections remain rare \citep{RN227, RN252}. A lower velocity limit for runaway stars has been suggested by \citet{RN137} at $\sim$5\,km\,s$^{-1}$, as stars ejected with this velocity can still travel a considerable distance in just a few Myr and end up tens of pc from any star-forming regions, satisfying distance-based runaway star definitions \citep[e.g.][]{RN190}. This subset of slower runaway stars has recently been termed walkaway stars \citep{RN136}.

Runaway and walkaway stars are thought to be the result of the same two ejection mechanisms. \citet{RN67} suggested that ejection of these stars is due to the binary supernova mechanism. This posits that in a close binary, the secondary star is ejected when the more massive primary reaches the end of its life and undergoes a core-collapse supernova. With a high enough kick velocity from the supernova, the main-sequence companion gets ejected due to binary disruption, almost always leaving an ejected singleton star. \citet{RN189} suggested an alternative mechanism due to dynamical interaction. In dense star-forming regions, stars interact with each other dynamically and close encounters between three or even four stars can lead to ejection of one or two of them. When a single, massive star interacts with a binary where the secondary is a lower-mass star, the single star can replace this secondary binary star, which is then ejected from the star cluster at a similar maximum velocity than in the binary supernova mechanism \citep{RN54}.

In this paper we use pure $N$-body simulations with differing initial conditions to investigate if the number and velocity distributions of unbound stars can allow us to place constraints on the initial density and velocity structure in star-forming regions. We aim to make predictions for observations of fast unbound stars from young star-forming regions that can be probed with \textit{Gaia} Data Release 2 (DR2). DR2 contains five-parameter astrometry (position, parallax and proper motion) for over 1.3 billion sources down to an apparent G-magnitude limit of G $\approx$ 21, whereas radial velocity information is only available for brighter sources ($\sim$7.2 million) \citep{RN238}. Our simulations provide us with 6D-parameter space results (position and velocity), but we focus on the 2D-plane and 2D-velocity, i.e. tangential velocity, which is calculated from proper motion and distance (or parallax) in observations. 

This paper is organised as follows. In section 2, we present the initial conditions used for the $N$-body simulations and our definition of unbound stars. Section 3 is dedicated to the results with a discussion of these results in section 4, followed by our conclusions in section 5. 

\section{Methods}

\subsection{Initial conditions}
Our simulated star-forming regions are set up with 1000 systems per simulation distributed across an initial radius of 1 pc. All systems are initially single stars (no primordial binaries) and their masses are randomly sampled for every single simulation from the \citet{RN203} initial mass function (IMF). This form of the IMF combines the Salpeter (\citeyear{RN204}) power-law slope for stars with masses above 1 M$_{\sun}$ with a Chabrier (\citeyear{RN200}) log-normal IMF approximation at the lower-mass end. The Maschberger IMF is described by a probability density function with three parameters $\alpha$ = 2.3 (power-law exponent for higher mass stars), $\beta$ = 1.4 (describing the IMF slope of lower-mass stars) and $\mu$ = 0.2 (average stellar mass) \citep{RN203}:
\begin{equation}
    p(m) \propto \cfrac{\left(\cfrac{m}{\mu}\right)^{-\alpha}}{\left(1+\left(\cfrac{m}{\mu}\right)^{1-\alpha}\right)^\beta}
    \label{eq:MaschbergerIMF}
\end{equation}

\noindent We sample stellar masses \textit{m} between 0.1 M$_{\sun}$ (we do not include brown dwarfs) and 50 M$_{\sun}$, resulting in total masses between $\sim$500-700 M$_{\sun}$ for each of our star-forming regions.

The spatial structure is set up following the method described in \citet{RN14}. It uses fractal distributions to define the observed substructure in young star-forming regions using a single parameter, the fractal dimension $D$. Starting with a cube with side $N_{\rm{div}}$ = 2, a parent particle is placed at the centre. This first parent cube is subdivided into equal-sized $N_{\rm{div}}^3$ sub-cubes with a first-generation descendant in each centre. Depending on the survival probability  $N_{\rm{div}}^{(D-\rm{3})}$ that is set by the fractal dimension $D$, these descendants can become parents themselves. For a low fractal dimension fewer descendants become parents, whereas more descendants survive when using a high fractal dimension. Descendants that do not survive are deleted along with their parent. The positions of the surviving particles are adjusted by adding a small amount of noise. This process continues until more stars than required are generated within the original cube. We cut a sphere from this cube and reduce the remaining stars down to the required number by random deletion. 

We use a set of four different fractal dimensions for our simulations to investigate a wide parameter space. Starting with highly substructured star-forming regions ($D$ = 1.6), we then gradually reduce the level of substructure ($D$ = 2.0 and $D$ = 2.6) finishing with a roughly uniform, smooth sphere ($D$ = 3.0).

Like the spatial structure, the velocity structure in our simulations is also set-up to mimic observed star-formation environments. Molecular gas clouds show turbulence that can be passed down to the stars that form from them. The velocity dispersion increases with the size of the clouds. In molecular clouds, large velocity dispersions can occur on large scales whereas on small scales there are smaller dispersions, i.e. similar velocities \citep{RN27}. Star formation occurs in filamentary structures within these gas clouds, where the velocity dispersion is low \citep{RN281}. To represent this velocity structure in our simulations we follow \citet{RN14}, which results in close stars with similar velocities and distant stars with different velocities. The process starts by assigning a random velocity to the parents. The next generation inherits this velocity, which is in turn adjusted by a random component that gets smaller with every following generation. The velocities of the stars are finally scaled to five different global virial ratios. The global virial ratio $\alpha_{\rm{vir}}$ describes the ratio of total kinetic energy $T$ of all stars to the modulus of the total potential energy $\varOmega$ of all stars,  $\alpha_{\rm{vir}}$ = $ T/|\varOmega|$. A star-forming region in virial equilibrium has a global virial ratio $\alpha_{\rm{vir}}$ = 0.5, with subvirial regions at values below and supervirial ones above. 

In our parameter space, we investigate star-forming regions initially in virial equilibrium as well as two regions that are initially subvirial ($\alpha_{\rm{vir}}$ = 0.1 and $\alpha_{\rm{vir}}$ = 0.3) and two supervirial ($\alpha_{\rm{vir}}$ = 1.0 and $\alpha_{\rm{vir}}$ = 1.5) initial settings. These global virial ratios describe the bulk motion of the stars as a whole. On local scales stars have similar, correlated velocities, meaning star-forming regions can be locally subvirial even if they are not subvirial on a global scale. This can lead to local, but not global collapse during the early dynamical evolution of the star-forming region \citep{RN4,RN1}.

We use the $N$-body integrator {\tt kira} from the {\tt Starlab} package \citep{RN236,RN193} to evolve our star-forming regions over a defined time period. The integrator uses an input $N$-body system defined by our initial conditions and evolves it over time giving output at different snapshots. The motion of the stars in the simulations is followed using a fourth-order, block-time-step Hermite scheme \citep{RN209}. 

With four different initial fractal dimensions and five different initial virial ratios, we run 20 simulations of each of the 20 combinations for a time period of 10 Myr to cover the early phases of the evolution of a star-forming region. The only changes within the simulations sharing the same initial conditions are the random number seed used to initiate the creation of the fractal (i.e. initial positions and velocities of stars) and the sampling of stellar masses from the IMF. For each set of initial conditions, we combine the results of all 20 simulations, thus creating a larger data set for analysis. Our star-forming regions do not have a gas potential and there is no external/tidal field applied. The stars do not undergo stellar evolution and are not in primordial binaries or initially mass-segregated. This allows us to identify the effects of different initial spatial and velocity substructure on the unbound population from young star-forming regions. In future work, we will include both stellar evolution and primordial binaries to quantify the effect each has on stars becoming unbound from these regions.

\subsection{Unbound stars and fractions by mass class}
We consider a star $i$ to be unbound once it has positive total energy (i.e. its kinetic energy $T_{i}$ is larger than the modulus of its potential energy $\varOmega_{i}$). Its kinetic energy is given by:
\begin{equation}
    T_{i} = \frac{1}{2} m_{i} |\mathbfit{v}_{i} - \mathbfit{v}_{cr}|^{2},
	\label{eq:kin_en}
\end{equation}
where $m_{i}$ is the mass of star $i$ and $\mathbfit{v}_{i}$ and $\mathbfit{v}_{cr}$ are the velocity vectors of this star and of the centre of the region, respectively. The potential energy of the star $i$ is given by the sum of the potential energy between star $i$ and every other star $j$:
\begin{equation}
    \mathit{\Omega_{i}} = - \sum_{i \neq j} \cfrac{G m_{i} m_{j}}{r_{ij}},
	\label{eq:pot_en}
\end{equation}
where $G$ is the gravitational constant, $m_{i}$ and $m_{j}$ are the stellar masses of $i$ and $j$ and $r_{ij}$ is the distance between them.
After identifying all unbound stars in each snapshot, we divide them up into two mass classes (MC): low/intermediate-mass (<8 M$_{\sun}$) and high-mass ($\geq$8 M$_{\sun}$) stars. We then calculate unbound fractions by normalising the number of unbound stars (UB) by the total number of stars (TOT) in that specific mass class: 
\begin{equation}\label{eq:Unbound fraction}
\text{Unbound fraction} = \cfrac{N_{MC,UB}}{N_{MC,TOT}}
\end{equation}
We estimate the standard error of the mean (SE) as a representation of the uncertainty connected to the unbound fractions, where $s$ is the sample standard deviation and $n$ is the number of simulations: 
\begin{equation}\label{eq:StandardError}
SE = \frac{s}{\sqrt{n}} 
\end{equation}

\begin{figure}
    \centering
	\includegraphics[width=0.95\linewidth]{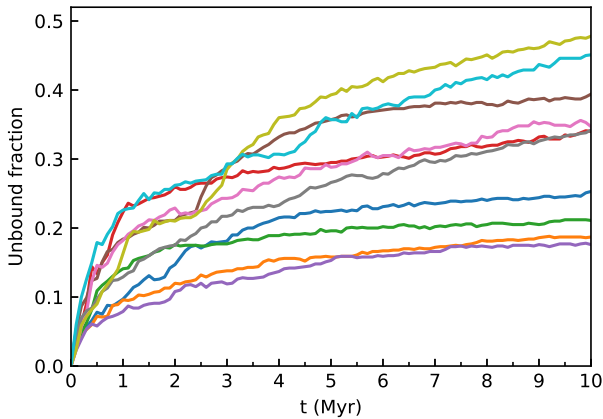}
    \caption{Unbound fractions from ten simulations (initially subvirial $\alpha_{\rm{vir}}$ = 0.3, with high level of initial substructure $D$ = 1.6) showing the spread of the unbound fractions between statistically identical simulations.}
    \label{fig:Ejected2}
\end{figure}

The uncertainty is caused by the stochastic nature of the underlying dynamical evolution \citep{RN4,RN5}. In our parameter space study, this different evolution is evident in the different unbound fractions from statistically identical, individual simulations as shown in Fig. \ref{fig:Ejected2}. This figure illustrates how different the unbound fractions can be for ten simulations with the same initial conditions (initially subvirial ($\alpha_{\rm{vir}}$ = 0.3) and high levels of substructure ($D$ = 1.6)). The different lines represent the fractions of unbound stars as a function of time and in this example they can increase over the simulation time to values between $\sim$18-48 per cent after 10 Myr.

\section{Results}
For the following analysis of velocities, we focus on 2D-velocities to allow us to make predictions for proper motion observations, such as from the recent \textit{Gaia} DR2 \citep{RN238}. In observations we have a fixed two-dimensional plane, whereas the choice of 2D-plane from simulations is arbitrary. The 2D-velocity results shown in this section represent the tangential velocity in the xy-plane (i.e. calculated as the motion across the sky would be in observations), however any other choice of 2D-plane gives us the same results after considering statistical noise.

\begin{figure*}
    \centering
	\includegraphics[width=\textwidth]{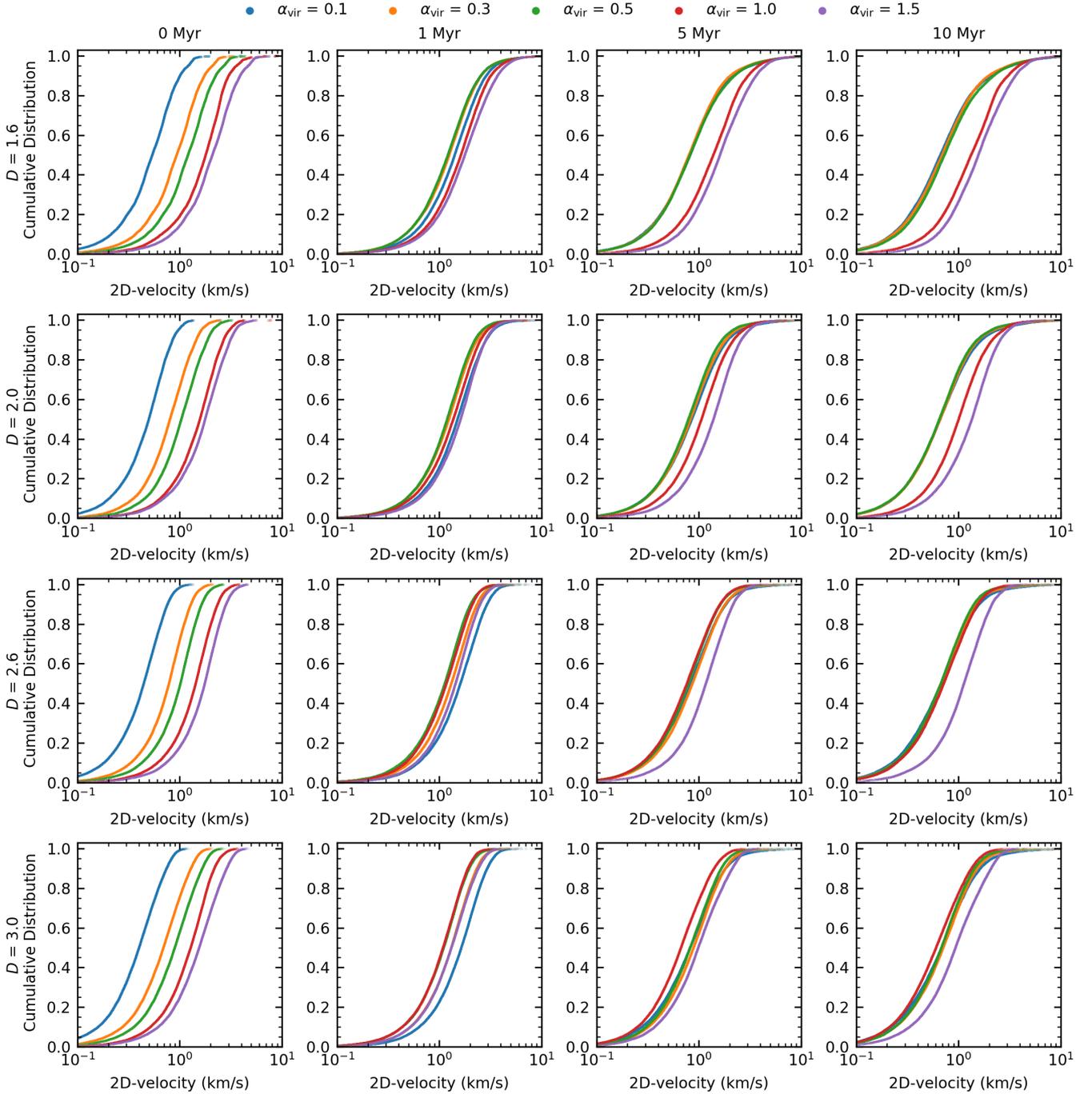}
    \caption{Cumulative 2D-velocity distributions at four different simulation times (in columns: 0 Myr, 1 Myr, 5 Myr, 10 Myr) and for the different initial condition sets. Each row represents a different fractal dimension from $D$ = 1.6 (top row) to $D$ = 3.0 (bottom row). The five different initial virial ratios ($\alpha_{\rm{vir}}$ = 0.1 (blue), $\alpha_{\rm{vir}}$ = 0.3 (orange), $\alpha_{\rm{vir}}$ = 0.5 (green), $\alpha_{\rm{vir}}$ = 1.0 (red), $\alpha_{\rm{vir}}$ = 1.5 (purple)) are shown in each panel for each fractal dimension and time.}
    \label{fig:Cumulative1}
\end{figure*}

\subsection{Cumulative 2D-velocity distributions of all stars}
We first focus on the cumulative distributions of the 2D-velocities and analyse how these evolve over the time period covered by our simulations. For each set of initial conditions, the cumulative distributions contain all stars from 20 simulations. In Fig. \ref{fig:Cumulative1}, we see the evolution of the cumulative distributions of the 2D-velocity at four different times from the left to the right column (0, 1, 5 and 10 Myr). From the top row to the bottom, we see that the four different fractal dimensions at 0 Myr show almost identical cumulative velocity distributions for all five initial virial ratios. This is to be expected as the virial ratio acts as a scaling factor for the initial velocities.

During the first 1 Myr, star-forming regions that are initially highly to moderately substructured ($D \leq$ 2.0) collapse and undergo violent relaxation \citep[e.g.][]{RN60,RN297,RN296,RN298,RN299,RN4,RN300} with subvirial regions ($\alpha_{\rm{vir}} <$ 0.5) collapsing rapidly to form bound, spherical clusters \citep{RN5}. Some of the initially virialised regions ($\alpha_{\rm{vir}}$ = 0.5) undergo a local collapse in regions of high substructure. Even though they are initially virialised on a global scale, they can be subvirial locally resulting in a localised collapse. Star-forming regions with little or no initial substructure ($D \geq$ 2.6) collapse only when they are also initially subvirial. 

At 1 Myr (second column), the velocity distributions of different initial virial ratios show similar velocities for identical levels of initial substructure. Initially highly subvirial regions ($\alpha_{\rm{vir}}$ = 0.1) that are slowest at the start of the simulations attain similar velocities to initially virialised and supervirial regions when $D \leq$ 2.0 or higher velocities when $D \geq$ 2.6. Violent relaxation leads to an increase in velocity, which is highest in highly subvirial, substructured initial conditions.

After 5 Myr and 10 Myr (third and fourth column), in initially more substructured regions ($D \leq$ 2.0) the evolution of the cumulative distributions follows a similar pattern.The bound, initially subvirial or virialised regions ($\alpha_{\rm{vir}} \leq$ 0.5) have very similar velocity distributions as the initially subvirial regions approach virial equilibrium after violent relaxation. Initially supervirial regions ($\alpha_{\rm{vir}} >$ 0.5) remain unbound and at higher average velocities. The difference between the subvirial/virial and supervirial distributions becomes clearer the older the simulated regions get, as the initially subvirial/virialised regions slow down compared to the initially supervirial ones.

Star-forming regions with less substructure initially ($D \geq$ 2.6) do not show the clear separation of velocity distributions between subvirial/virial and supervirial initial ratios. Only initially highly supervirial regions ($\alpha_{\rm{vir}}$ = 1.5) have a velocity distribution at later times that can be distinguished from those with lower virial ratios. The initially smooth, sphere-like regions ($D$ = 3.0) still show a grouping together of the velocity distributions after 5 Myr. The two initially supervirial distributions  ($\alpha_{\rm{vir}}$ = 1.0 and 1.5) are located either side of the initially subvirial and virialised ones. Despite both being supervirial, they exhibit considerably different velocity distributions. Moderately supervirial regions ($\alpha_{\rm{vir}}$ = 1.0) have the slowest, whereas highly supervirial regions ($\alpha_{\rm{vir}}$ = 1.5) have the fastest cumulative 2D-velocities. This behaviour continues for the remaining 5 Myr and at the end of our simulations the moderately supervirial cases are still indistinguishable from those of initially subvirial/virialised ($\alpha_{\rm{vir}} \leq$ 0.5) cases.

\begin{figure*}
    \centering
	\includegraphics[width=\textwidth]{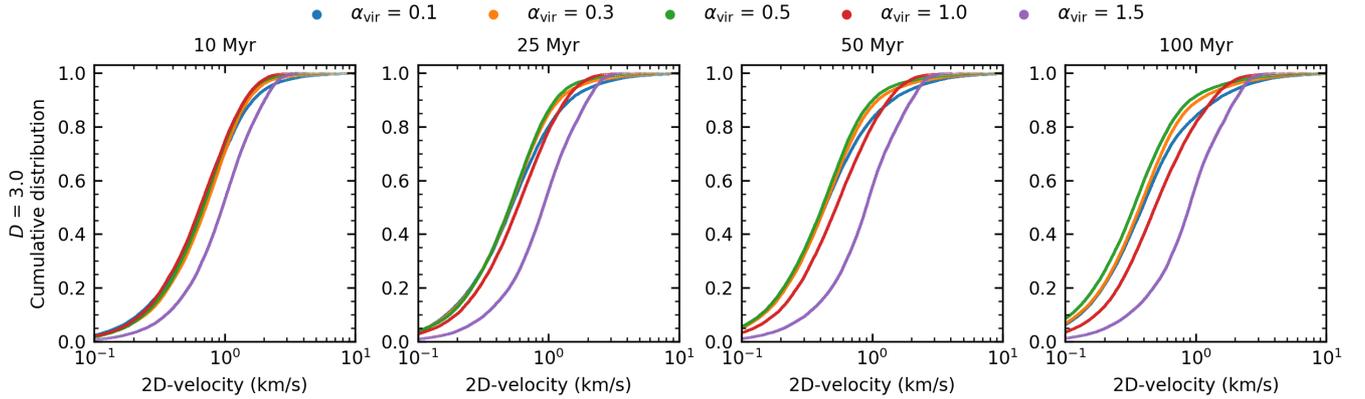}
    \caption{Long-term evolution of the cumulative 2D-velocity distributions at four different simulation times (10, 25, 50 and 100 Myr) for the five different initial virial ratios ($\alpha_{\rm{vir}}$ = 0.1 (blue), $\alpha_{\rm{vir}}$ = 0.3 (orange), $\alpha_{\rm{vir}}$ = 0.5 (green), $\alpha_{\rm{vir}}$ = 1.0 (red), $\alpha_{\rm{vir}}$ = 1.5 (purple)) and constant fractal dimension ($D$ = 3.0).}
    \label{fig:Long-term_evolution}
\end{figure*}

\subsubsection{Long-term evolution of initially smooth star-forming regions}
For these initially smooth star-forming regions ($D$ = 3.0), we follow the evolution of their cumulative distributions for a longer time period. We evaluate if they evolve differently or just more slowly than initially more substructured star-forming regions. The evolution of these smooth regions is shown at 10 Myr, 25 Myr, 50 Myr and 100 Myr in Fig. \ref{fig:Long-term_evolution}. 

The cumulative distributions for initially subvirial and virialised regions ($\alpha_{\rm{vir}} \leq$ 0.5) continue to be similar as they are in a state of virial equilibrium. The velocity distribution for the moderately supervirial regions ($\alpha_{\rm{vir}}$ = 1.0, red) starts to become distinguishable from the initially subvirial/virialised regions after 50 Myr, as these regions slow down compared to the moderately supervirial one. 

But even after 100 Myr the velocities of moderately supervirial regions are still much closer to those of initially subvirial/virial star-forming regions than the highly supervirial scenario. Initially smooth, supervirial star-forming regions appear to evolve in a similar fashion than the more substructured regions but on a much longer timescale. The long-term evolution of the cumulative distributions shows that the average velocities decrease at later times for initially subvirial/virialised regions, as the global gravitational field of the bound clusters cause stars to decelerate.

\subsection{Unbound fractions of stars from initially subvirial and virialised regions}
In this section we turn to unbound fractions for initially subvirial and virialised star-forming regions ($\alpha_{\rm{vir}} \leq$ 0.5). We exclude the two supervirial scenarios as in these globally unbound, expanding regions most stars are born unbound. In our simulations, we do not have any stellar evolution, so stars can only become unbound due to dynamical interactions with other stars \citep{RN189} and not from supernova kicks \citep{RN67}. In the absence of an external tidal field, lower-mass stars mainly become unbound due to effects of 2-body relaxation \citep{RN161} whereas high-mass stars require dynamical interactions with other high-mass stars in binaries or higher order multiple systems (e.g. trapezium-like) to become unbound \citep{RN38,RN17}. 

\begin{figure*}
    \centering
	\includegraphics[width=\textwidth]{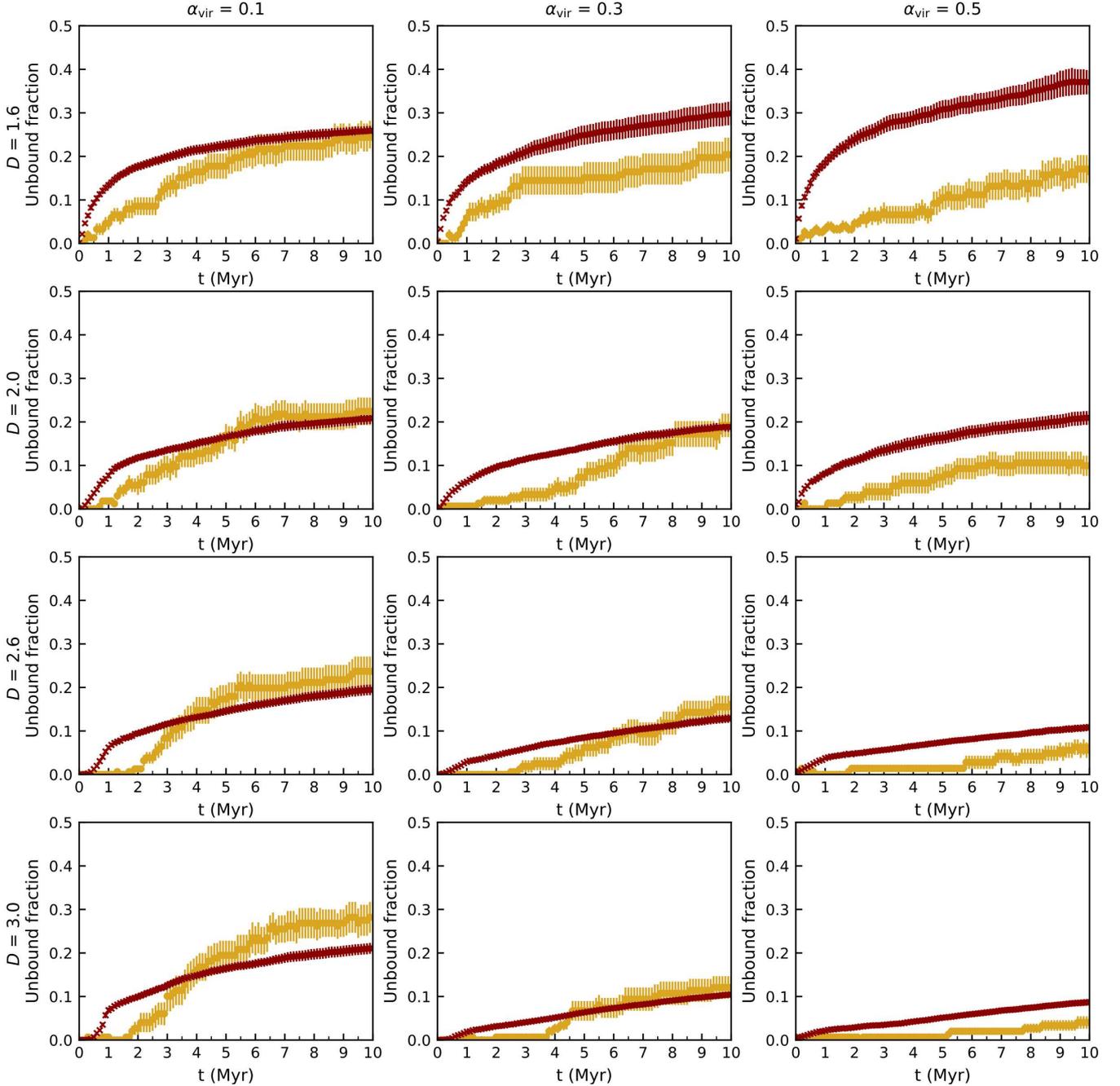}
    \caption{Unbound fractions by mass class for initially subvirial and virialised star-forming regions ($\alpha_{\rm{vir}} \leq$ 0.5). Each row represents a different fractal dimension starting from $D$ = 1.6 (top row) to $D$ = 3.0 (bottom row). The columns show the three subvirial and virial initial ratios. The red points represent the unbound fraction of low/intermediate-mass stars (<8 M$_{\sun}$) over the simulation time, whereas the yellow points represent the unbound fraction of high-mass stars (>8 M$_{\sun}$). The uncertainties of the fractions are calculated using the standard error of the mean (Eq. \ref{eq:StandardError}).}
    \label{fig:Ejected1}
\end{figure*}

\subsubsection{Effects of different levels of substructure in regions with the same initial virial ratio}
In Fig. \ref{fig:Ejected1}, unbound fractions for star-forming regions with an initially highly subvirial ratio ($\alpha_{\rm{vir}}$ = 0.1) are shown in the first column. With high levels of initial substructure ($D$ = 1.6, first row) stars in both mass classes show similar unbound fractions from 5 Myr to the end of the simulations. These regions, regardless of initial degree of substructure, will undergo rapid collapse and violent relaxation. While low/intermediate-mass stars become unbound early in the simulations, high-mass stars show a more gradual increase and match the lower-mass unbound fraction at $\sim$5 Myr. The lower-mass unbound fraction decreases with less initial substructure and settles on the same level of $\sim$20 per cent for more moderate amounts of initial substructure ($D$ = 2.0-3.0) after 10 Myr. At the end of our simulations, the high-mass unbound fractions in the four initial substructure scenarios reach final values between 22$\,\pm\,$3 per cent and 28$\,\pm\,$4 per cent. 

We see a delay in high-mass stars becoming unbound that increases the lower the level of initial substructure (i.e. higher fractal dimension $D$) in initially highly subvirial regions ($\alpha_{\rm{vir}}$ = 0.1, first column). In these simulations the degree of collapse reduces with lower amounts of initial substructure, resulting in a longer formation time for multiple star systems that can eject massive stars. The low/intermediate-mass stars also show a delay in stars becoming unbound for $D$ = 2.6-3.0. The delay is most obvious in regions with no initial substructure ($D$ = 3.0, bottom row). On average, only 7 stars (all are low/intermediate-mass) per simulation become unbound in the first $\sim$0.5 Myr. The lack of initial substructure combined with the low initial virial ratio appears to result in a ``balanced" collapse that keeps virtually all stars in a bound configuration for a short period of time ($\sim$0.5 Myr). 

In initially moderately subvirial simulations ($\alpha_{\rm{vir}}$ = 0.3, second column) the star-forming regions undergo an initial collapse but the degree of collapse is lower when compared to the highly subvirial simulations. We decrease the level of initial substructure and see a significant decrease in the low/intermediate-mass unbound fraction for every change in fractal dimension. After 10 Myr, the high-mass unbound fractions only slightly decrease (from 20$\,\pm\,$4 per cent to 19$\,\pm\,$3 per cent) for regions with initially high or moderate levels of substructure ($D \leq$ 2.0). Further decreasing the initial substructure reduces the high-mass unbound fraction to 16$\,\pm\,$2 per cent ($D$ = 2.6) and 12$\,\pm\,$3 per cent ($D$ = 3.0). The high-mass unbound fractions are only different for simulations with higher ($D \leq$ 2.0) and no initial substructure ($D$ = 3.0). 

In regions with initial fractal dimensions $D$ = 2.0-3.0, high-mass stars do not become unbound early in the simulations. The collapse happens fastest in initially highly substructured star-forming regions ($D$ = 1.6) and high-mass stars can become unbound much earlier than in less substructured star-forming regions. The lower the level of initial substructure, the longer it takes to form dynamical multiples that can eject high-mass stars \citep{RN38}. Our simulations suggest that it can take over 3 Myr longer for high-mass stars to become unbound when there is a lack of initial substructure in moderately subvirial initial conditions (lower, middle panels).  

In all simulations that are initially virialised ($\alpha_{\rm{vir}}$ = 0.5, third column), regardless of substructure, the unbound fraction of low/intermediate-mass stars is at least double the fraction of unbound high-mass stars after 10 Myr, which reaches $16\,\pm\,3$ per cent in initially highly substructured regions ($D$ = 1.6). In these star-forming regions 37$\,\pm\,$3 per cent of all low/intermediate-mass stars become unbound at the end of our simulations. 

Initially virialised, highly substructured star-forming regions can collapse locally and binary clusters can form \citep{RN210}. Binary clusters are a pairing of star clusters that are physically close to each other in space \citep{RN278,RN288,RN287,RN290,RN291}. They can be a result of the dynamical evolution of young star-forming regions as shown by \citet{RN210}. We see these binary clusters at the end of more than half of the 20 simulations and they appear to have an effect on the unbound fractions. The presence of binary clusters lowers the sub-cluster potential energy, effectively creating two smaller clusters with smaller potential wells. In consequence, stars require lower kinetic energy to become unbound. This increases the low/intermediate-mass unbound fraction, but does not affect the high-mass unbound fraction in the same way. Due to the form of the IMF, there is a much smaller number of high-mass stars present in our simulations. During the localised collapse into binary clusters, these high-mass stars can move to different local regions, reducing the likelihood of creating dynamical multiple systems that can eject high-mass stars from our regions.

We do find a higher unbound fraction for high-mass stars than low/intermediate-mass stars in several simulations with initially virialised, highly substructured conditions ($\alpha_{\rm{vir}}$ = 0.5, $D$ = 1.6) that do not result in the creation of binary clusters. In individual simulations where binary clusters are present, we see a higher than average fraction ($\sim$30 per cent compared to the average value of $\sim$16 per cent) of unbound high-mass stars when the low/intermediate-mass unbound fraction is high as well ($\sim$40-70 per cent) or when the absolute number of high-mass stars is high to begin with (i.e. 9 or more high-mass stars per simulation). This increases the chances of forming high-mass dynamical multiple systems, which would lead to more ejections.

Lower levels of initial substructure or smooth regions ($D$ = 2.0-3.0) that are initially virialised ($\alpha_{\rm{vir}}$ = 0.5) do not form binary clusters \citep{RN210}. In our simulations, this considerably reduces the unbound fractions. Star-forming regions that are initially in virial equilibrium and smooth ($D$ = 3.0) undergo very little dynamical evolution and most of the stars ($\sim$87 per cent) remain bound throughout the simulations. 

\subsubsection{Effects of different initial virial ratios in regions with the same levels of substructure}
For star-forming regions with a high degree of initial substructure ($D$ = 1.6, first row in Fig. \ref{fig:Ejected1}) increasing the initial virial ratio has the opposite effect on the unbound fractions in the two mass classes. The increase in initial kinetic energy (higher virial ratio) in the regions decreases the fraction of unbound high-mass stars whereas it increases the fraction of low/intermediate-mass unbound stars. While an initially highly subvirial region ($\alpha_{\rm{vir}}$ = 0.1) has the same unbound fraction after 10 Myr in both mass classes, the more virialised a highly substructured region is initially the higher its unbound fraction of low/intermediate-mass stars and the lower its high-mass unbound fraction. The low/intermediate-mass unbound fraction is highest in initially virialised regions due to the presence of binary clusters.

In regions with a lower level of initial substructure ($D$ = 2.0, second row) differences in initial virial ratio have no effect on the low/intermediate mass unbound fractions, which are virtually the same for all three initial virial ratio scenarios (values between 19$\,\pm\,$1 per cent and 21$\,\pm\,$2 per cent) at 10 Myr. The high-mass unbound fraction is highest in the initially most subvirial regions ($\alpha_{\rm{vir}}$ = 0.1). The degree of collapse is highest here and unstable multiple star systems can form quickly. After about 6 Myr, the high-mass unbound fraction reaches 21$\,\pm\,$3 per cent and starts to level out (22$\,\pm\,$3 per cent at 10 Myr), suggesting that unstable multiple star systems are no longer present or do not lead to any further high-mass star ejections. The initially more moderate, subvirial ($\alpha_{\rm{vir}}$ = 0.3) simulations have a similar high-mass unbound fraction than the virialised case in the first $\sim$6 Myr of the simulations (10$\,\pm\,$2 per cent vs. 9$\,\pm\,$2 per cent). The difference in initial virial ratio ($\alpha_{\rm{vir}}$ = 0.3 vs. 0.5) appears to have no effect on the early evolution of these simulated regions. Later in the simulation, the initially moderately subvirial ($\alpha_{\rm{vir}}$ = 0.3) regions continue to eject high-mass stars and reach an unbound fraction of 19$\,\pm\,$3 per cent after 10 Myr, which is a similar value than in the highly subvirial case ($\alpha_{\rm{vir}}$ = 0.1), whereas the high-mass unbound fraction in initially virialised regions levels out after $\sim$ 7 Myr and remains at 10$\,\pm\,$2 per cent to the end of the simulations at 10 Myr.

At low levels of or no initial substructure ($D$ = 2.6 and 3.0, third and fourth row) the low/intermediate-mass unbound fractions are highest when the regions are initially highly subvirial ($\alpha_{\rm{vir}}$ = 0.1) as these regions collapse initially. Even though the moderately subvirial ($\alpha_{\rm{vir}}$ = 0.3) regions initially collapse, this does not result in a higher low/intermediate unbound fraction than in the initially virialised regions that do not undergo collapse. When there is little or no initial substructure, star-forming regions will only collapse when the initial virial ratio is subvirial. The collapse increases the likelihood that unstable multiple systems form, which facilitates the ejection of high-mass stars. With higher initial virial ratios, these multiple systems take longer to form or do not form at all. As a result, high-mass stars take longer to become unbound and the final unbound fractions at 10 Myr are lower the more virialised and smooth the regions are initially.

\subsection{2D-velocity of unbound stars from initially subvirial and virialised star-forming regions}

\subsubsection{Cumulative 2D-velocity distributions}

\begin{figure}
    \centering
	\includegraphics[width=0.85\linewidth]{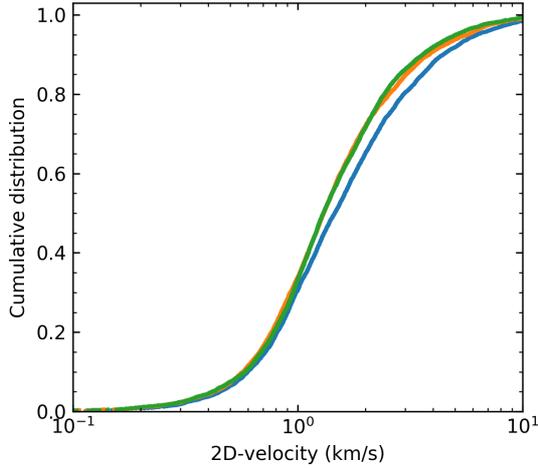}
    \caption{Cumulative distributions for unbound stars showing the 2D-velocities at 10 Myr with initial fractal dimension $D$ = 1.6 for initially subvirial and virialised clusters. The distributions for $\alpha_{\rm{vir}}$ = 0.1 (blue), $\alpha_{\rm{vir}}$ = 0.3 (orange) and $\alpha_{\rm{vir}}$ = 0.5 (green) are shown zoomed in to the central part of the curve, highlighting that for these initial conditions, the velocities of the unbound stars do not differ much between different virial ratios for the same degree of initial spatial and kinematic substructure.}
    \label{fig:Cumulative2}
\end{figure}

In Fig. \ref{fig:Cumulative2}, we show the 2D-velocity cumulative distributions for unbound stars from initially subvirial/virialised regions ($\alpha_{\rm{vir}} \leq$ 0.5) with a high level of initial substructure ($D$ = 1.6) at 10 Myr. As we have seen for all (bound and unbound) stars in Fig. \ref{fig:Cumulative1} the cumulative distributions in initially subvirial/virialised simulations are very similar for all four initial fractal dimensions. The cumulative distributions of unbound stars in Fig. \ref{fig:Cumulative2} show a similar picture of very similar distributions for a fractal dimension of $D$ =  1.6 (initially highly substructured regions). Even for these initially substructured regions where we see a more dynamic early evolution (i.e. violent relaxation and initial collapse), it is difficult to distinguish between different initial virial ratios at the end of the simulations.

\begin{figure}
    \centering
	\includegraphics[width=0.85\linewidth]{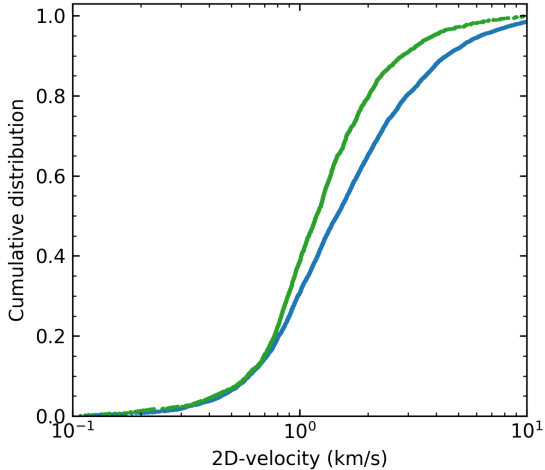}
    \caption{Cumulative distributions for unbound stars showing the 2D-velocities at 10 Myr for an initially highly substructured and subvirial region with $D$ = 1.6 and $\alpha_{\rm{vir}}$ = 0.1 (blue) and an initially almost smooth and virialised region with $D$ = 2.6 and $\alpha_{\rm{vir}}$ = 0.5 (green). The comparison illustrates that the more substructured and subvirial a star-forming region is initially, the faster the unbound stars escape.}
    \label{fig:Cumulative3}
\end{figure}

\citet{RN3} analysed the spatial and velocity distributions of very different initial conditions after 4 Myr with a smaller, but similar, set of initial conditions. He found that the cumulative velocity distributions differ between the initial conditions and that the initially moderately subvirial, substructured simulations result in higher velocity unbound stars compared with initially virialised simulations with a low level of substructure. Fig. \ref{fig:Cumulative3} illustrates the cumulative velocity distributions of very different initial conditions after 10 Myr: initially highly substructured and highly subvirial simulations ($D$ = 1.6, $\alpha_{\rm{vir}}$ = 0.1, blue) compared to simulations with a low level of substructure that are initially virialised ($D$ = 2.6, $\alpha_{\rm{vir}}$ = 0.5, green). We also find that unbound stars from substructured, subvirial regions are moving at higher 2D-velocities (after 10 Myr) however the differences between the distributions are not quite as large as in \citet{RN3}. This highlights that cumulative velocity distributions can only distinguish between vastly different initial spatial and velocity conditions.

\begin{figure*}
    \centering
	\includegraphics[width=\textwidth]{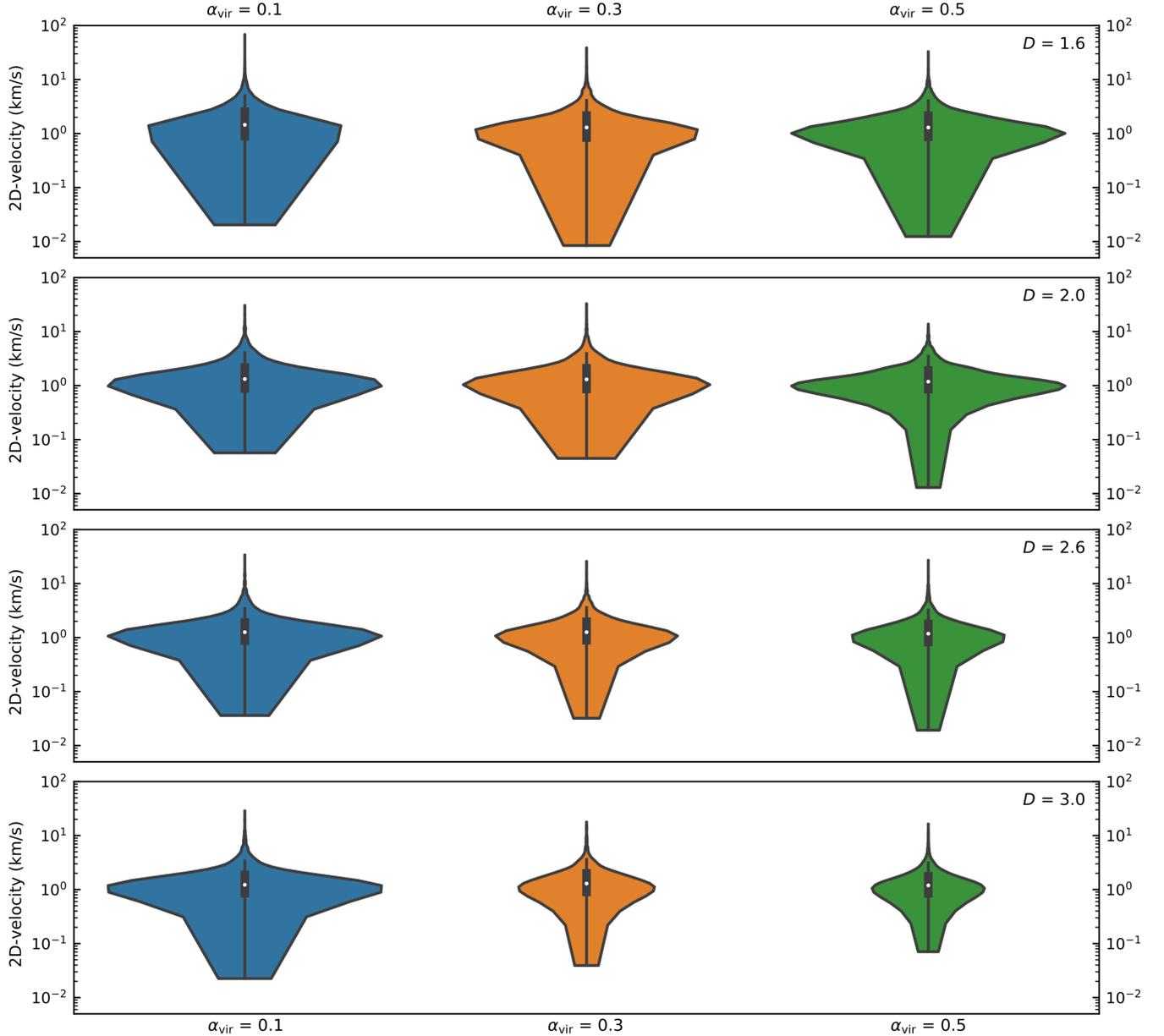} 
    \caption{Violin plots showing the 2D-velocity distributions of unbound stars at 10 Myr from all initially subvirial and virialised regions ($\alpha_{\rm{vir}}$ = 0.1 (blue), $\alpha_{\rm{vir}}$ = 0.3 (orange)) and virialised ($\alpha_{\rm{vir}}$ = 0.5 (green)). The violins are scaled by count, the wider the violins are at any point the more stars in our regions have this 2D-velocity. The larger the violins overall, the more stars have become unbound during the simulation time. The thick vertical bar in the centre shows the interquartile range with the white dot representing the median. The long vertical line represents the 95 per cent confidence interval.}
    \label{fig:Violin}
\end{figure*}

\subsubsection{Violin plots of 2D-velocity distributions}
Violin plots are a data visualisation technique that combines a box plot with a density trace or a kernel density estimate \citep{RN207}. Like box plots, violin plots also show the median and interquartile range for a variable, as well as any asymmetries and outlier data. They can be useful when comparing distributions of a variable (2D-velocity) over different categories (initial conditions for star-forming regions). Unlike box plots, violin plots includes all data from the underlying distribution and give information about the shape of the distribution. They show all peaks and the position of those peaks, their amplitude, and give insight into any presence of clustering in the data. The outer shape represents all data, with the widest parts corresponding to the value (i.e. 2D-velocity) with the highest probability of occurring in the population \citep{RN207}, which can be also interpreted as the most common 2D-velocity in our case.

Fig. \ref{fig:Violin} shows the 2D-velocity distributions on a log-scale for all initially subvirial and virialised regions (left to right) and all four fractal dimensions (decreasing degree of initial substructure - top to bottom) after 10 Myr. The plots include all unbound stars from 20 simulations combined and represent an average. The wider each of the violin plots is at any point, the more stars are likely to have this 2D-velocity. For each fractal dimension (in each row), the width of the violin plot is scaled by the total number of unbound stars for this initial virial ratio. For two violin plots with the same total number of unbound stars, the widest part will have the same width. However the absolute number of stars with this velocity can be different, e.g. for fractal dimension $D$ = 2.0 (second row) the blue ($\alpha_{\rm{vir}}$ = 0.1) and green ($\alpha_{\rm{vir}}$ = 0.5) violin plots both contain a total of $\sim$4100 unbound stars from 20 simulations each, resulting in the widest part of their violin plots having the same width in Fig. \ref{fig:Violin}. Due to difference in the distributions, there are $\sim$80 more stars at the most common velocity for the initially virialised (green) violin plot.

The thick vertical bar in the centre represents the interquartile range with the white dot representing the median. The thin long vertical line represents the 95 confidence interval. We use a bandwidth following the \citet{RN289} reference rule to smooth our data for the violin plots\footnote{https://seaborn.pydata.org/generated/seaborn.violinplot.html}. The violin plots are cut at the low-velocity end and only show the actual data points there, instead of the tails of the underlying Gaussian kernel density estimate. This allows us to identify the lowest actual 2D-velocity directly from the plot and avoids the appearance of negative 2D-velocities.

\begin{figure*}
    \centering
    \begin{minipage}[t]{1.0\columnwidth}
        \centering
        \vspace{0pt}
    	\includegraphics[width=0.95\linewidth]{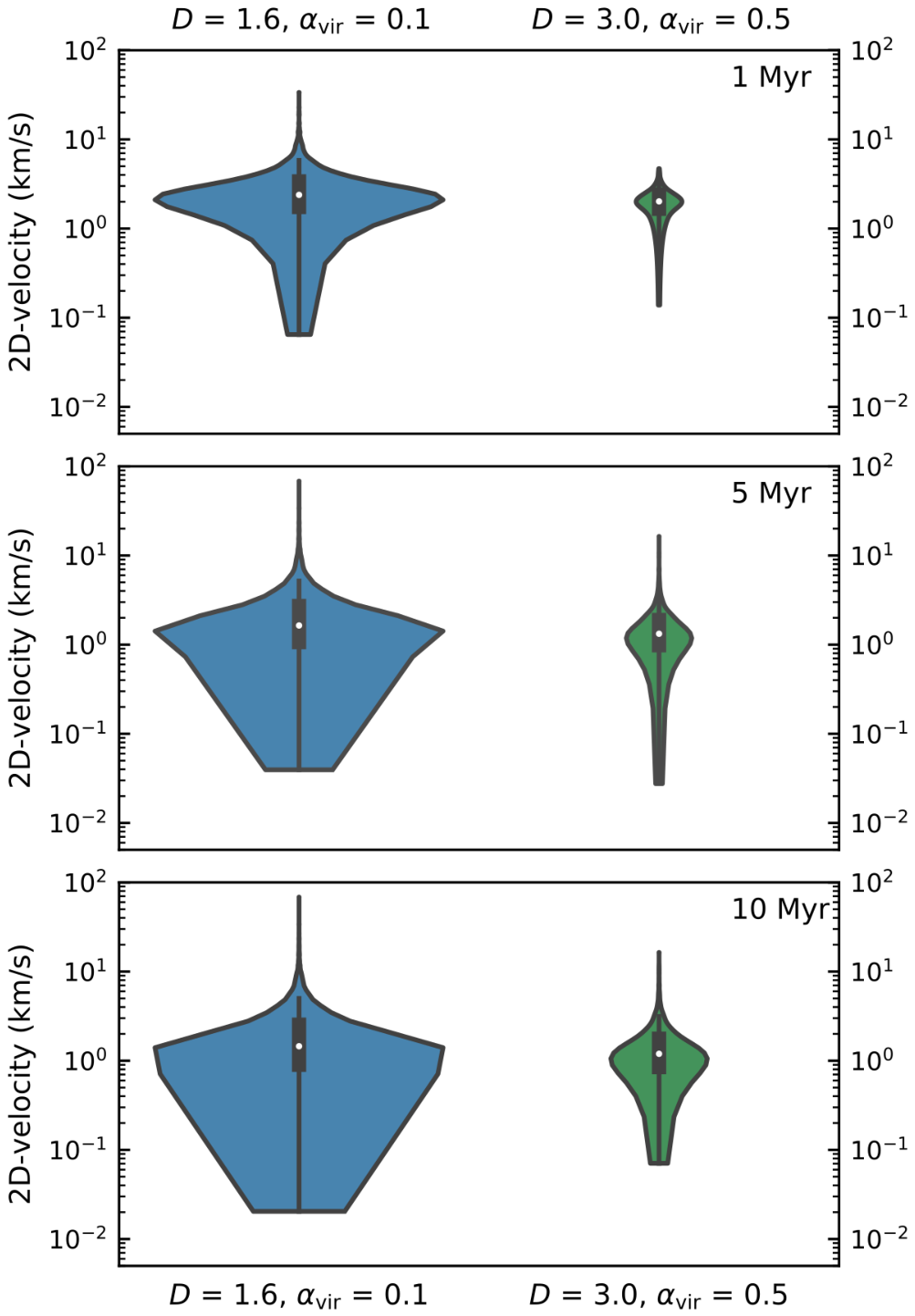}
        \caption{Violin plots showing 2D-velocity (xy-plane) distributions of unbound stars at three simulation times for two selected initial conditions (initially subvirial, substructured (blue) and initially virialised with no substructure (green)).}
        \label{fig:Violin_2D}
    \end{minipage}
    \hfill{}
    \begin{minipage}[t]{1.0\columnwidth}
        \centering
        \vspace{0pt}
        \includegraphics[width=0.95\linewidth]{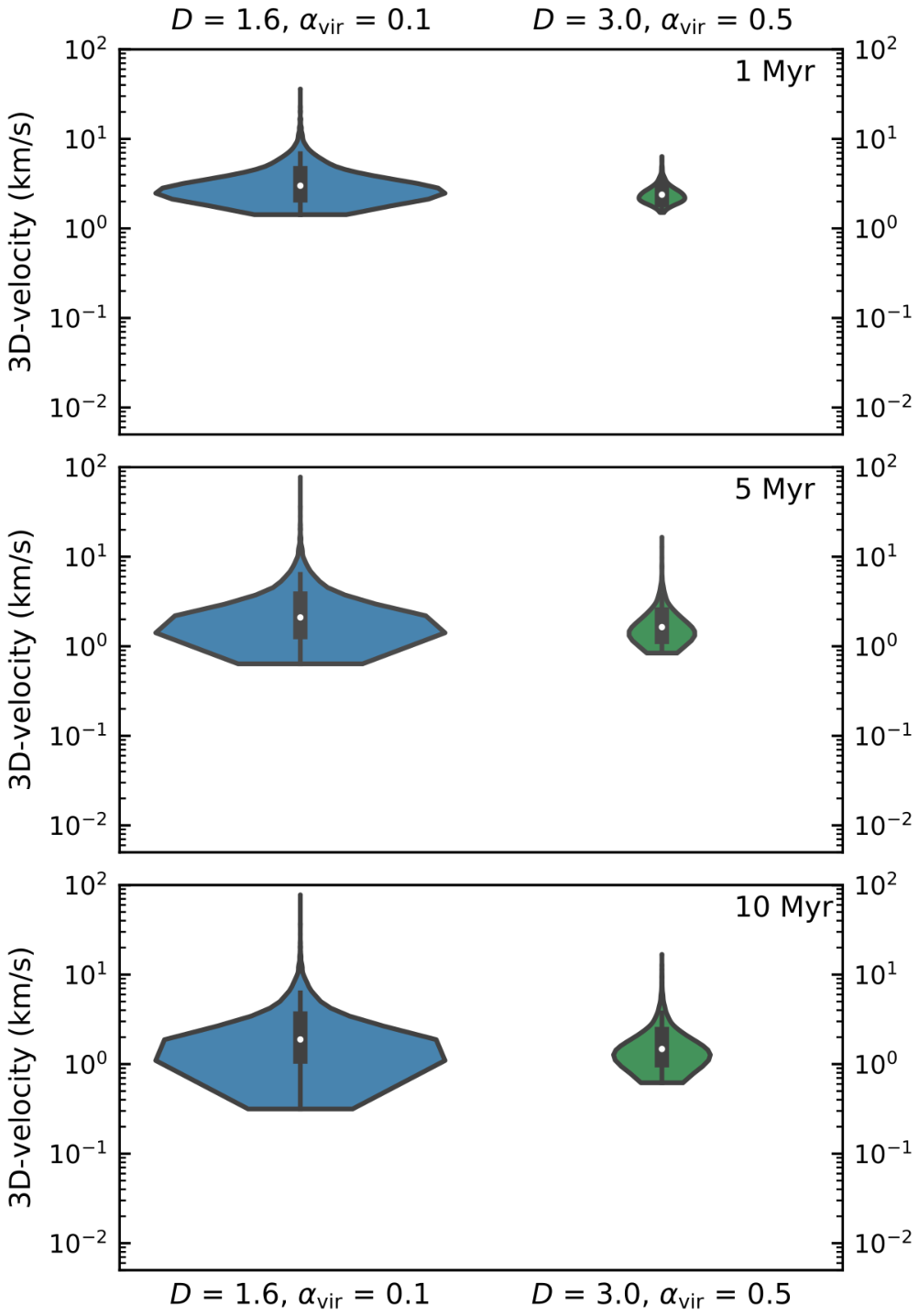}
        \caption{Violin plots showing 3D-velocity distributions of unbound stars at three simulation times for two selected initial conditions (initially subvirial, substructured (blue) and initially virialised with no substructure (green)).}
        \label{fig:Violin_3D}
    \end{minipage}
\end{figure*}

Initially highly substructured regions ($D$ = 1.6, Fig. \ref{fig:Violin} first row) have a large number of unbound stars for all three initial virial ratios. The fastest stars are ejected from initially highly subvirial regions ($\alpha_{\rm{vir}}$ = 0.1, blue) with the peak velocity reaching $\sim$70\,km\,s$^{-1}$. These regions have fewer unbound stars ($\sim$260 per simulation) in total and fewer stars at similar velocities with a wider spread of velocities around $\sim$1\,km\,s$^{-1}$ compared to the two higher virial ratio scenarios. Despite these differences, the median velocity is similar ($\sim$1.5\,km\,s$^{-1}$) to the other two scenarios ($\sim$1.3\,km\,s$^{-1}$ - both for $\alpha_{\rm{vir}}$ = 0.3 and 0.5). A large number of unbound stars from highly substructured, moderately subvirial regions ($\alpha_{\rm{vir}}$ = 0.3, orange) move at a similar 2D-velocity of $\sim$1\,km\,s$^{-1}$ after 10 Myr, creating noticeable arms in the violin plots. The total number of unbound stars increases to $\sim$300 per simulation. The arms become most pronounced in the initially virialised case ($\alpha_{\rm{vir}}$ = 0.5, green) with $\sim$370 unbound stars per simulation. Despite the increase in the total number of unbound stars, the most common velocity remains around $\sim$1\,km\,s$^{-1}$. The higher the initial virial ratio in initially highly substructured regions, the more likely it is that unbound stars are moving with more similar velocities, whereas unbound stars are more evenly spread over different velocities in initially more subvirial regions. 

With a lower level of initial substructure ($D$ = 2.0, second row) the shape of the distributions changes for all three initial virial ratios. The shape of the velocity distributions of the two initially subvirial scenarios ($\alpha_{\rm{vir}}$ < 0.5) is now almost identical. The violin plot for highly subvirial regions is wider than the moderately subvirial scenario, meaning more stars become unbound (20 more stars per simulation). We see the pronounced arms in the violin plots now also for highly subvirial regions with less spread in the 2D-velocities. The fastest stars from the two subvirial regions now only reach $\sim$30\,km\,s$^{-1}$ and their median velocities are almost identical ($\sim$1.3\,km\,s$^{-1}$). In initially virialised regions ($\alpha_{\rm{vir}}$ = 0.5, green), the arms in the 2D-velocity become even more pronounced at $\sim$1\,km\,s$^{-1}$. The maximum velocity is lower ($\sim$13\,km\,s$^{-1}$) than in the subvirial cases ($\sim$30\,km\,s$^{-1}$), however the median velocity remains similar ($\sim$1.2\,km\,s$^{-1}$).

In regions with little or no initial substructure ($D \geq$ 2.6, third and fourth row), initially highly subvirial regions ($\alpha_{\rm{vir}}$ = 0.1) show a similar violin shape to the more substructured regions ($D$ = 2.0) and also have a similar number of unbound stars ($\sim$200-210 unbound stars per simulation). The sizes of the violins shrink considerably (i.e. fewer unbound stars) for initially moderately subvirial ($\alpha_{\rm{vir}}$ = 0.3) and virialised ($\alpha_{\rm{vir}}$ = 0.5) regions and we see $\sim$90-130 unbound stars per simulation. This indicates a much less dynamical early evolution with the number of unbound stars only $\sim$30 per cent of what they are in simulations with the highest level of initial substructure. Despite this, the violins retain their overall familiar shape of having arms around the most common velocity of $\sim$1\,km\,s$^{-1}$ and a median velocity ($\sim$1.2-1.3\,km\,s$^{-1}$), which is similar to all other initial conditions.

\begin{figure*}
    \centering
	\includegraphics[width=\textwidth]{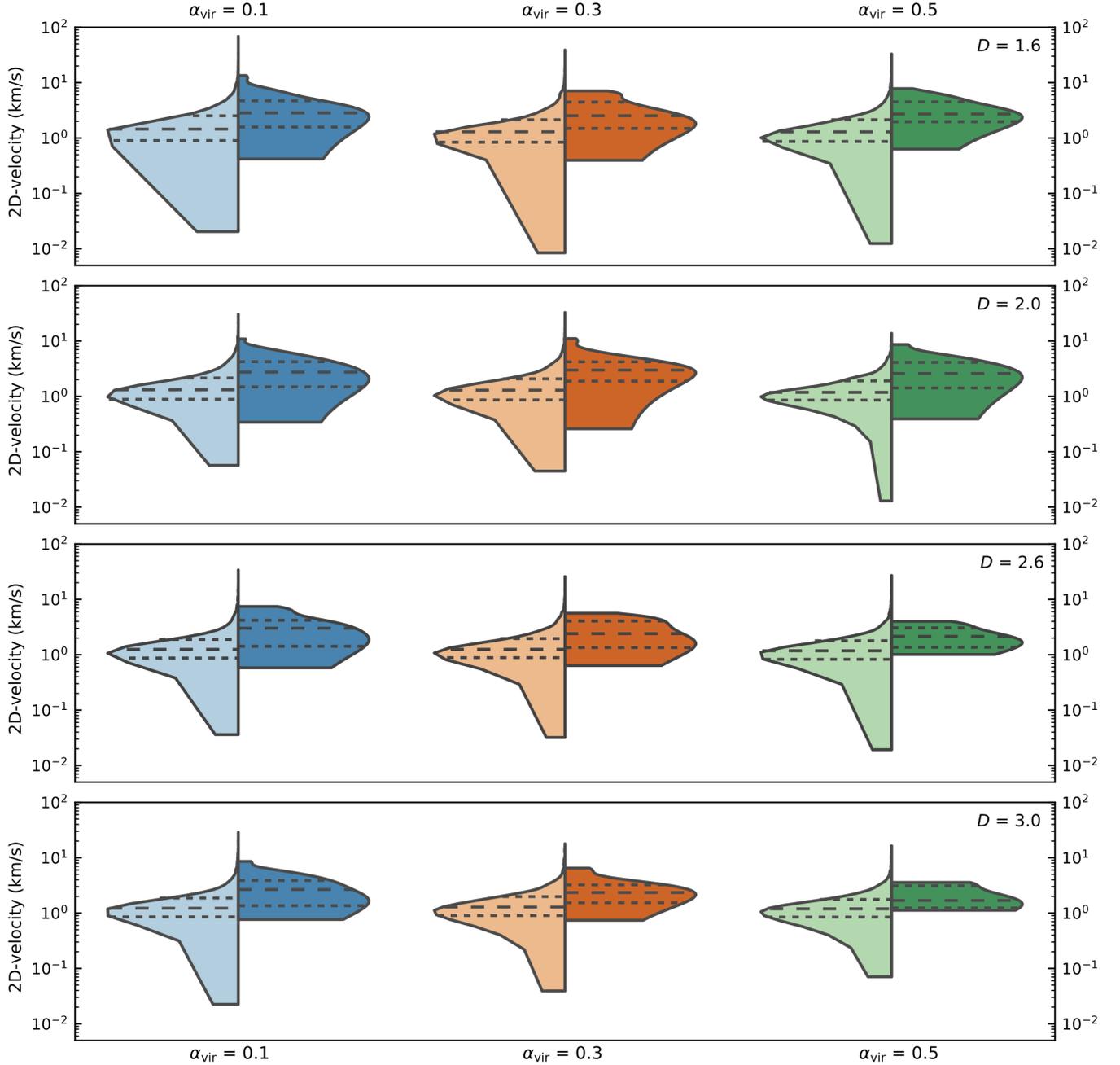} 
    \caption{Violin plots showing the 2D-velocity distributions of unbound stars at 10 Myr split by mass class (low/intermediate-mass - left half, high-mass - right h) from all initially subvirial and virialised clusters ($\alpha_{\rm{vir}}$ = 0.1 (blue), $\alpha_{\rm{vir}}$ = 0.3 (orange)) and virialised ($\alpha_{\rm{vir}}$ = 0.5 (green)). All plots are scaled to have the same width as there is only a very small number of unbound high-mass stars. The widest part of the each violin half represents the 2D-velocity with the highest probability. Dashed lines represent the median and the interquartile range, the 95 per cent confidence interval is no longer shown.}
    \label{fig:Violin_mass}
\end{figure*}

Our unbound definition is based on stars reaching the escape velocity (total positive energy) from the star-forming regions, which is $\sim$3\,km\,s$^{-1}$ in our simulated regions. In Fig. \ref{fig:Violin} we see that the minimum 2D-velocity of unbound stars can be as low as $\sim$0.03\,km\,s$^{-1}$ after 10 Myr. Once unbound stars leave the denser parts of a star-forming region, they interact with fewer or no other stars and slow down gradually. However, the apparent slow-down in our simulations by up to two orders of magnitude is likely due to projection effects. Fig. \ref{fig:Violin_2D} shows violin plots for two, very different initial conditions (blue - initially highly subvirial, substructured and green - initially virialised, no substructure) at three different times during the simulations. Already after 1 Myr, a low-velocity tail forms in 2D-space that extends to velocities an order of magnitude lower than the escape velocity. In full 3D-velocity space in Fig. \ref{fig:Violin_3D}, we see that after 1 Myr the lowest velocities are only 1-2\,km\,s$^{-1}$ lower than the escape velocity. This suggests that unbound stars slow down, however not to the extent suggested by the 2D-velocities.  

\begin{figure*}
    \centering
	\includegraphics[width=0.95\textwidth]{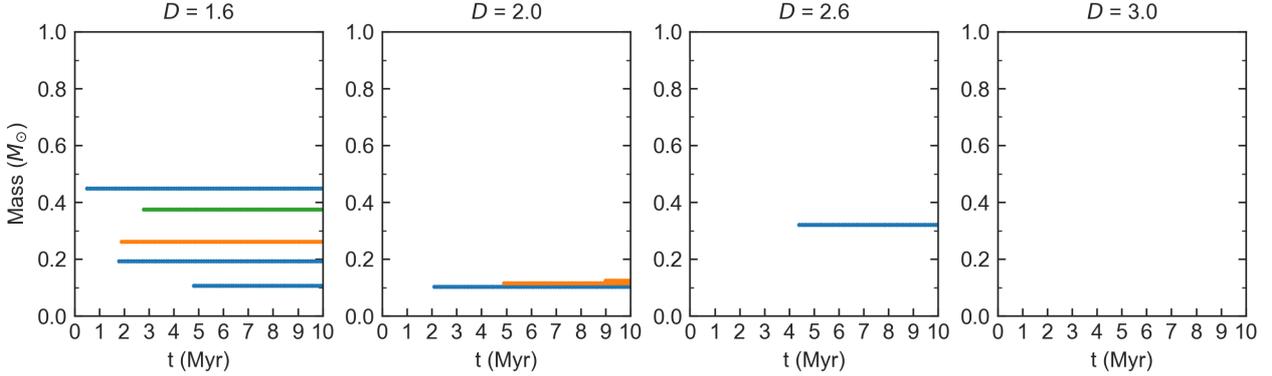}
    \caption{Runaway stars (2D-velocity > 30\,km\,s$^{-1}$) by mass over the simulation time for the four fractal dimensions and the three different initial virial ratios: subvirial ($\alpha_{\rm{vir}}$ = 0.1 (blue) and $\alpha_{\rm{vir}}$ = 0.3 (orange)) and virialised ($\alpha_{\rm{vir}}$ = 0.5 (green). The y-axis is limited to 1 M$_{\sun}$, as all of our runaway stars have very low mass.}
    \label{fig:Runaway}
\end{figure*}

This 2D-projection effect could affect cluster membership identification when observing proper motion (or 1D-radial velocity) in isolation. Depending on relative position to the cluster, these ``slow" unbound stars could be identified as not having originated from the cluster at all due to being too far away or still bound due to their central location in the star-forming region. However, our simulations suggest that only $\sim$1 per cent of these unbound stars with low 2D-velocities are located in the central parts of star-forming regions after 10 Myr. This limits the extent of mistakenly assigning membership to ``slow" unbound stars, when only proper motion information is available.

In Fig. \ref{fig:Violin_mass} we use split violin plots to show the 2D-velocities separately for the two mass-classes. The plots are now scaled to the same width as we have at most $\sim$40 unbound high-mass stars compared with over 7000 lower-mass unbound stars from a set of 20 simulations. The widest part of each half still represents the 2D-velocity with the highest probability of occurring. Dashed lines represent the median and the interquartile range, the 95 per cent confidence interval is no longer identified on the plots. The violin plots are again cut at the low-velocity end and only show the actual data points, instead of the tails of the underlying Gaussian kernel density estimate. This allows us to identify the lowest actual 2D-velocity directly from the plot and avoids the appearance of negative 2D-velocities.

In Fig. \ref{fig:Violin_mass} we see that the shape of the low/intermediate-mass violins is nearly identical to the shape of the total population of unbound stars in Fig. \ref{fig:Violin} as most unbound stars are lower mass. Due to the low number of unbound high-mass stars the velocity distributions of unbound high-mass stars can have a jagged outline depending on the bandwidth used. We use the same bandwidth setting \citep[following][]{RN289} as in Fig. \ref{fig:Violin} resulting in the right half (unbound high-mass stars) of our split violin plots in Fig. \ref{fig:Violin_mass} appearing as a smooth distribution despite the small sample size. A small sample size can make conclusions from violin plots unreliable and we limit our interpretation of them to general differences in median, minimum and maximum velocity between the two mass-classes. To gain more insight into the velocity distributions of unbound high-mass stars using violin plots would require an increase in the sample size, i.e. a much higher number of simulations.

For all initial condition scenarios at 10 Myr, high-mass unbound stars have a higher median (and interquartile range) than the low/intermediate-mass stars and also a much higher minimum 2D-velocity. The mechanism for high-mass stars to become unbound is different to that of low/intermediate-mass stars. High-mass stars will only become unbound from our star-forming regions after a dynamical interaction with other massive stars in multiples. These dynamical interactions make unbound high-mass stars move faster on average, however the fastest stars are in fact from the low-mass end. The differences in 2D-velocities between the mass classes is present in all initial condition combinations, so is not affected by the initial spatial or velocity structure in the star-forming regions.

\subsection{Runaway and walkaway stars}
Finally, we analyse how effective star-forming regions with different initial conditions are at ejecting runaway and walkaway stars. We only use 2D-velocity and the lower boundary value of 30\,km\,s$^{-1}$ \citep[e.g.][]{RN255, RN276, RN190, RN137} for our runaway definition and velocities between 5-30\,km\,s$^{-1}$ for walkaways \citep{RN137, RN136}. 

Fig. \ref{fig:Runaway} shows all stars from 20 simulations per initial condition moving with a 2D-velocity (xy-plane) above 30\,km\,s$^{-1}$. All of them are from the low end of the mass spectrum, not a single runaway star is more massive than 0.5 M$_{\sun}$. We have the highest number of runaway stars from initially highly substructured, subvirial regions ($\alpha_{\rm{vir}}$ = 0.1, $D$ = 1.6) regardless of the choice of 2D-plane. Only the fastest one is present in all three 2D-planes and is moving with a 2D-velocity between 50-70\,km\,s$^{-1}$ depending on the choice of plane. The other two runaways have lower velocities between 30-40\,km\,s$^{-1}$. With at most three ejected runaways from a set of 20 simulations, we see that regardless of the initial velocity or spatial structure, runaway stars are rare from our chosen initial conditions. 

\begin{figure*}
    \centering
    \begin{minipage}[t]{0.95\textwidth}
        \centering
        \includegraphics[width=1.0\linewidth]{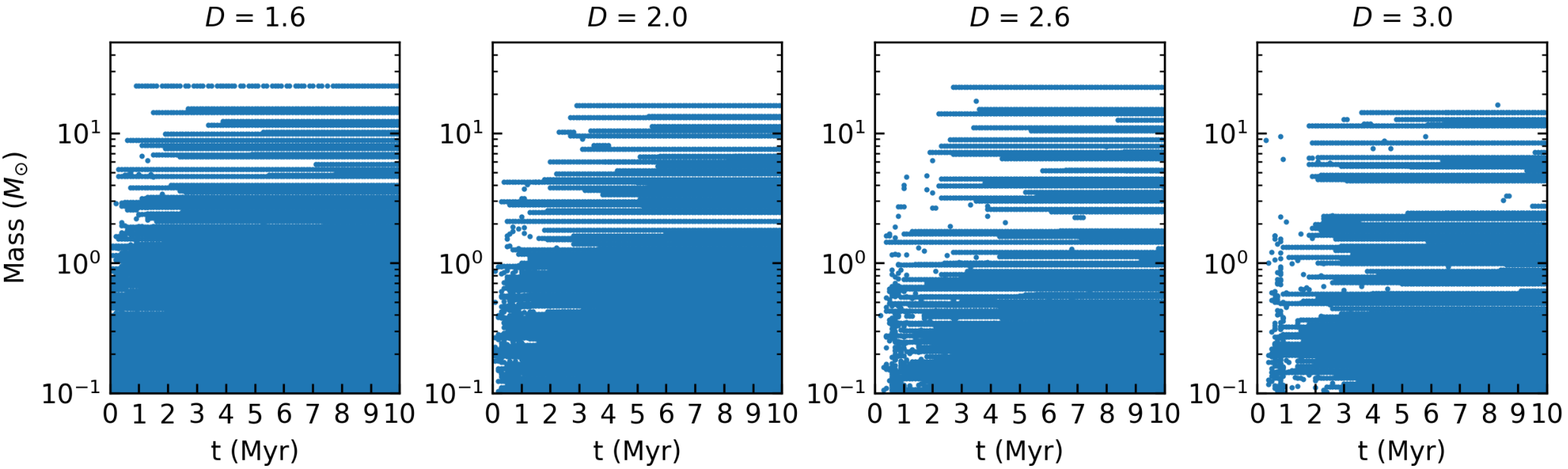}
    \end{minipage}
    \begin{minipage}[t]{0.95\textwidth}
        \centering
    	\includegraphics[width=1.0\linewidth]{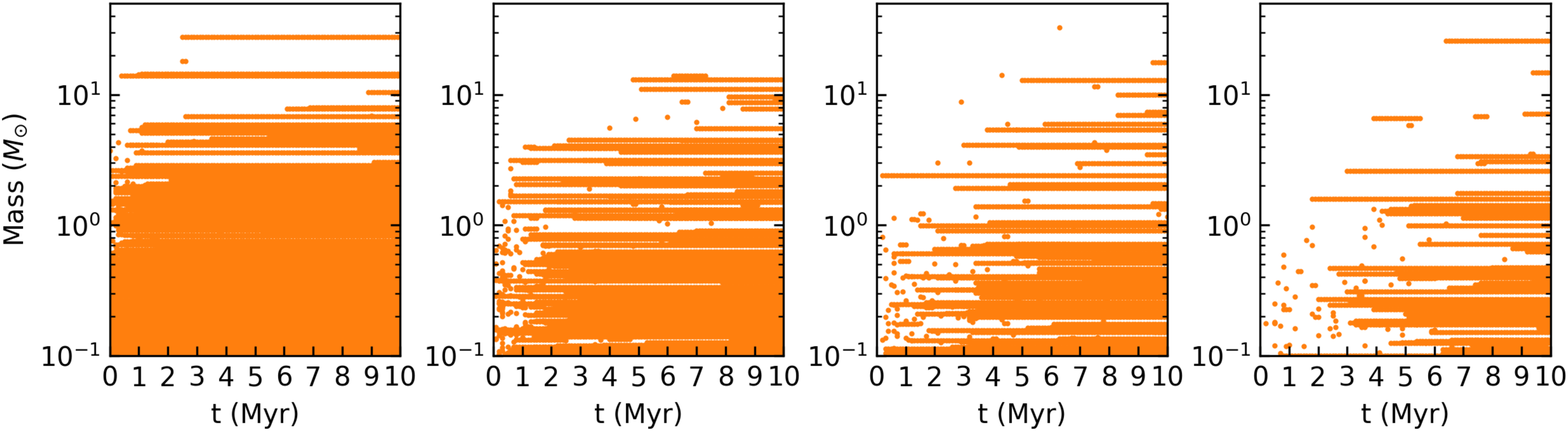}
    \end{minipage}
    \begin{minipage}[t]{0.95\textwidth}
        \centering
    	\includegraphics[width=1.0\linewidth]{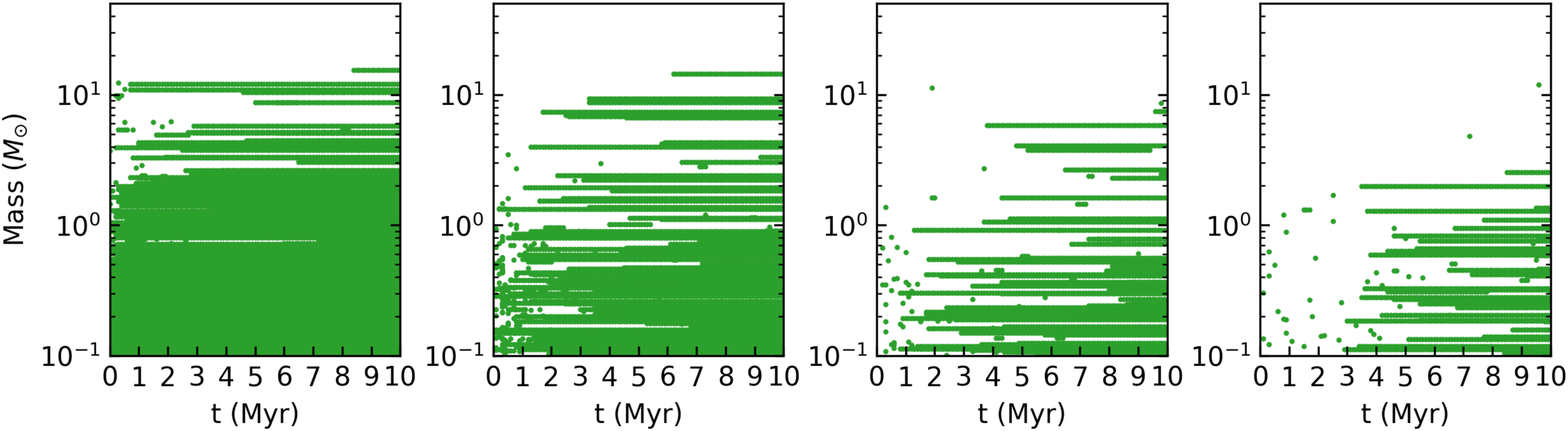}
    \end{minipage}
    \caption{Walkaway stars (2D-velocity: 5-30\,km\,s$^{-1}$) by mass (using a log-scale) over the simulation time for the four fractal dimensions and the three different initial virial ratios: subvirial ($\alpha_{\rm{vir}}$ = 0.1 (blue, top row) and $\alpha_{\rm{vir}}$ = 0.3 (orange, middle row)) and virialised ($\alpha_{\rm{vir}}$ = 0.5 (green, bottom row)). A few stars (single points) are only identified as walkaways for a few snapshots, this is due to them being ejected close to the lower walkaway velocity boundary and slowing down to fall below the boundary shortly after ejection.}
    \label{fig:Walkaway}
\end{figure*}

Going to walkaway velocities (5-30\,km\,s$^{-1}$) produces a few high-mass walkaways and a large number of low-mass walkaways across all initial conditions. Fig. \ref{fig:Walkaway} shows all walkaways from the 20 simulations across each initial condition set. The more violent the early evolution of a star-forming region is, the higher the number of walkaway stars. In the most violently evolving initial condition set-up - initially highly substructured ($D$ = 1.6) and highly subvirial ($\alpha_{\rm{vir}}$ = 0.1), we have on average $\sim$0.5 high-mass walkaways per simulation and $\sim$20 low/intermediate-mass walkaways per simulation. 

The lower the initial level of substructure (larger fractal dimension $D$) the lower the overall number of walkaway stars, with initially more subvirial regions (Fig. \ref{fig:Walkaway} top row) producing more walkaway stars, which are also ejected earlier in the simulations. We see a number of temporary walkaways that appear as walkaways only for a few snapshots. These are stars ejected just at the minimum walkaway velocity. After ejection they slow down and disappear from our plots once they drop below 5\,km\,s$^{-1}$ (minimum walkaway velocity), however this does not mean that they have been recaptured by the star-forming region. Initially virialised star-forming regions with no substructure ($\alpha_{\rm{vir}}$ = 0.5, $D$ = 3.0 - bottom right panel) produce on average only 2 low/intermediate-mass walkaways per simulation. This is an order of magnitude fewer walkaway stars than in the initial condition scenario (initially highly substructured ($D$ = 1.6), highly subvirial ($\alpha_{\rm{vir}}$ = 0.1) - top left panel) that produces the largest number of walkaway stars.

\section{Discussion}

We summarise the results of our $N$-body simulations as follows. Cumulative velocity distributions of star-forming regions with different initial conditions have limited usefulness in clearly distinguishing between different initial spatial and velocity structure. When comparing the long-term evolution of regions with different levels of initial substructure, regions with high levels of initial substructure evolve very quickly kinematically, with supervirial regions (unbound by definition) showing the fastest 2D-velocities. The cumulative velocity distributions of unbound stars from initially subvirial and virialised simulations are difficult to distinguish after 10 Myr and only show differences for extremely different initial conditions (see Fig. \ref{fig:Cumulative3}).

The unbound fraction differs considerably for different combinations of initial spatial and velocity structure. This suggests that the unbound population around young, bound star clusters could possibly be used to draw conclusions about the initial conditions. Around initially smooth ($D$ = 3.0), virialised ($\alpha_{\rm{vir}}$ = 0.5) star-forming regions, we find a low number of ejected stars (slow walkaways, but no runaways) and virtually no unbound high-mass stars  after 10 Myr. Around initially substructured, subvirial regions that have undergone violent relaxation, we find a large number of unbound low/intermediate-mass stars. We also find a few high-mass ejected stars (at walkaway velocities) and one low-mass runaway star in three of the 20 simulations. The unbound fractions are possibly influenced by our choice of initial density as higher densities increase the likelihood of encountering and interacting with other stars. 

Initial densities can differ greatly from those currently observed due to the amount of dynamical evolution that a region undergoes. The level of spatial substructure in a region can constrain the dynamical evolution of regions with different initial densities - the higher the initial density, the quicker substructure is erased \citep{RN8}. Our simulated star-forming regions have been set up with a high, median local initial density (10$^{3}$-10$^{4}$ M$_{\sun}$ pc$^{-3}$). 

After about 1 Myr, regions with initial spatial substructure have evolved into smooth, centrally concentrated regions, whose densities can be directly compared to observed star-forming regions. The density in our simulations after a few Myr is 10$^{1}$-10$^{3}$ M$_{\sun}$ pc$^{-3}$ \citep{RN8} comparable to many nearby star-forming regions where observed present-day densities do not exceed $\sim$400 M$_{\sun}$ pc$^{-3}$ \citep[e.g.][]{RN277}.

High-mass stars are less likely to become unbound than low/intermediate-mass stars if a region is not initially very subvirial. When they do escape from their birth environment they do so at higher velocity and become at least walkaway stars (> 5\,km\,s$^{-1}$). With our chosen initial conditions, high-mass stars do not reach the velocity regime of runaway stars. Only the evolution of star-forming regions that are initially subvirial ($\alpha_{\rm{vir}}$ < 0.5) and/or substructured ($D \leq$ 2.0) is dynamic enough to produce any runaway stars, all of which are low-mass. This is in apparent contrast to observations, where due to observational bias, predominantly high-mass runaways are found \citep{RN263} as they are more luminous and easier to observe. Historically, the definition of runaway stars is based on OB stars \citep{RN67}, following \citet{RN263}, we suggest to extend this definition to lower-mass stars. Lower-mass stars appear to reach runaway velocities more often than higher-mass stars and these could be found around many young star-forming regions when testing our predictions with \textit{Gaia} DR2. 

Using data from \textit{Gaia} DR1 \citep{GaiaDR1}, \citet{RN294} report that two of the most massive stars (HD46223 and HD46106) in NGC2244 are moving away from each other and from the centre of this young cluster at a larger velocity than the other cluster stars. They suggest that HD46223 has been ejected from the cluster, possibly due to dynamical interactions with other massive stars in the centre. The inferred velocity of 1.38\,km\,s$^{-1}$ from its proper motion \citep{RN294} is far below the lower velocity boundary for walkway stars and it is unclear if this star is actually unbound. Our simulated star-forming regions (1000 single stars) have an escape velocity of $\sim$3\,km\,s$^{-1}$. NGC2244 is estimated to have $\sim$2000 members \citep{RN295} suggesting that HD46223 might not have reached escape velocity and might still be bound to the cluster despite its apparent ejection. In our simulations we also see massive stars moving outwards after dynamical interactions at velocities higher than their surroundings. If they are moving more slowly than the escape velocity they will remain bound to the cluster, slow down and eventually return in direction of the cluster centre.

Violin plots show that the velocity distributions do indeed differ between initial conditions, particularly when the regions are initially highly substructured. These distributions also indicate that the vast majority of low/intermediate-mass stars become unbound at just around the escape velocity. We show that 2D-velocity information appears to be an underestimate of the full 3D-velocity for a proportion of unbound stars. This can have implications for membership determination of young star-forming regions, where full velocity parameter space information is not available. The \textit{Gaia} DR2 data set contains a much larger number of stars only with proper motion data, missing information about the radial velocity for many fainter stars. If the 2D-velocity is indeed an underestimate of the full space velocity for some stars, we might mistakenly assign a cluster membership to stars with slow proper motions or not be able to trace back stars to their birth cluster.   

Escaping, ejected or unbound stars from simulations have been studied previously \citep[e.g.][]{RN239,RN46,RN3,RN235,RN241,RN222}. \citet{RN3} found a similar connection between unbound stars (i.e. number, velocity, spatial distribution) and the initial substructure and virial ratio with a more limited set of initial conditions. Other studies \citep{RN239,RN241} used Plummer spheres \citep{RN198} to set up the initial spatial distribution of the clusters and included primordial binaries. The conclusion from these studies was that the number and mass fraction of unbound stars depend strongly on the initial cluster radius or initial density and to a lesser extent on the parameters of the primordial binaries \citep{RN239, RN241} or the initial virial ratio \citep{RN239}. With their inclusion of primordial binaries, the results of these studies are not directly comparable to ours. 

Our results show that differences in the initial spatial substructure can have a considerable effect on the fraction, the velocity and the masses of unbound stars. Due to the lack of stellar evolution in our short simulation time of 10 Myr, we miss the effects of supernova kicks causing stars to become unbound due to the binary supernova ejection mechanism \citep{RN67}. In our simulations, binaries will only form dynamically (i.e. are not present from the beginning of our simulations) and we may therefore be underestimating the impact of the dynamic ejection mechanism \citep{RN189} as we only find a few lower-mass runaways stars. \citet{RN235} showed that a higher fraction of primordial binaries increases the number of higher-velocity (20-100\,km\,s$^{-1}$) stars.

\section{Conclusions}

In this paper, we present $N$-body simulations of star-forming regions set up with a range of different initial spatial and velocity structures. We investigate if the dynamical evolution results in differences in the unbound population after 10 Myr. The conclusions from our simulations are summarised as follows.

\renewcommand{\labelenumi}{(\roman{enumi})}
\begin{enumerate}
  \item Cumulative 2D-velocity distributions of all stars in simulated star-forming regions cannot provide strong insights into the long-term-term evolution of star-forming regions with differing initial spatial and velocity structure. When focussing on unbound stars, clear differences in the cumulative distributions are only found when comparing vastly different initial conditions.
  \item Unbound fractions of stars of different masses show clear differences between the initial conditions and could prove useful to distinguish between initial spatial and velocity structures. Only when a region is initially very subvirial can we expect to find a higher fraction of unbound high-mass stars than low/intermediate-mass stars in the vicinity of the region.
  \item If high-mass stars manage to escape their birth region, they are likely to reach at least walkaway velocities. However based on our simulations, not every young star-forming region will create a high-mass runaway or walkaway star.   
  \item Most low/intermediate-mass stars leave the regions at velocities just above the escape velocity. However, the fastest stars from our simulations are also low/intermediate-mass stars. We see a number of low/intermediate-mass walkaway stars from every initial condition set. This number increases for regions that evolve more dynamically (more initial substructure and lower virial ratio). As a result, we should find at least a small number of these stars around virtually every young and high-density star-forming region. The fact that most observed fast stars are still high-mass is very likely due to observational bias/limitations. This changes with \textit{Gaia} DR2 where five-parameter space astrometry for stars down to sub-solar mass is already available for star-forming regions nearby \citep{RN238}. We will use this data to search for walkaway and runaway stars from young star-forming regions in a future paper. 
\end{enumerate}

\section*{Acknowledgements}
CS acknowledges PhD funding from the 4IR STFC Centre for Doctoral Training in Data Intensive Science and thanks ESTEC/ESA in Noordwijk, in particular the \textit{Gaia} team, for hosting her as a visiting researcher for six months. RJP acknowledges support from the Royal Society in the form of a Dorothy Hodgkin Fellowship.



\bibliographystyle{mnras}
\bibliography{Main_document} 







\bsp	
\label{lastpage}
\end{document}